\documentclass[aps,prc,superscriptaddress,twoside,twocolumn,nofootinbib,10pt,%
showpacs,floatfix]{revtex4-1}

\usepackage{amsmath,amssymb}
\usepackage{graphicx,bm}
\usepackage{slashed}
\usepackage{epstopdf}
\usepackage{ulem} 
\usepackage[usenames]{color}
\usepackage{subfigure}
\usepackage{float}

\begin{document}

\title{The mass of heavy-light mesons in a constituent quark picture with partially restored chiral symmetry }

\author{Aaron Park}
\email{aaron.park@yonsei.ac.kr}
\affiliation{Department of Physics and Institute of Physics and Applied Physics,
Yonsei University, Seoul 120-749, Korea}

\author{Philipp Gubler}
\email{pgubler@riken.jp}
\affiliation{ECT{$^\ast$}, Villa Tambosi, 38123 Villazzano (Trento), Italy}

\author{Masayasu Harada}
\email{harada@hken.phys.nagoya-u.ac.jp}
\affiliation{Department of Physics,  Nagoya University, Nagoya, 464-8602, Japan}

\author{Su Houng Lee}
\email{suhoung@yonsei.ac.kr}
\affiliation{Department of Physics and Institute of Physics and Applied Physics,
Yonsei University, Seoul 120-749, Korea}

\author{Chiho Nonaka}
\email{nonaka@hken.phys.nagoya-u.ac.jp}
\affiliation{Kobayashi-Maskawa Institute for the Origin of Particles and the Universe,  Nagoya University, Nagoya, 464-8602, Japan}
\affiliation{Department of Physics,  Nagoya University, Nagoya, 464-8602, Japan}

\author{Woosung Park}
\email{diracdelta@hanmail.net}
\affiliation{Department of Physics and Institute of Physics and Applied Physics,
Yonsei University, Seoul 120-749, Korea}

\date{\today}

\begin{abstract}
We probe effects of the partial chiral symmetry restoration to the mass of heavy-light mesons in a constituent quark model by changing the constituent quark mass of the light quark. 
Due to the competing effect between the quark mass and the linearly rising potential, whose contribution to the energy increases as the quark mass decreases, the heavy-light meson 
mass has a minimum value near the constituent quark mass typically used in the vacuum.  
Hence, the meson mass increases as one decreases the constituent quark mass consistent with recent QCD sum rule analyses, which show an increasing $D$ meson mass as the 
chiral order parameter decreases.  
\end{abstract}
\maketitle

\section{Introduction}

Investigating the relation between chiral symmetry breaking and physical observables has been the subject of great interest up to this date 
as these effects can be probed in 
heavy ion collisions and/or nuclear target experiments. 
The restoration of chiral symmetry has been linked to vector meson masses, $\sigma$ meson masses, the quenching of $f_\pi$ and even to the $\eta'$ 
mass  (see, e.g. Refs.~\cite{Hatsuda:1986gu,Hatsuda:1991ez,Brown:1991kk,%
Rapp:1999ej,Harada:2000kb,Hayano:2008vn,Leupold:2009kz,Lee:2013es,GublerOhtani:2014}).

Of particular interest is the heavy-light quark meson  as it embodies both chiral symmetry breaking and heavy quark symmetry.  Recently, one of us in collaboration with other authors 
employed QCD sum rules to show that the $D$ meson mass increases as the chiral symmetry is partially restored \cite{Suzuki:2015est}. 
A qualitatively similar conclusion was obtained in Ref.~\cite{Hilger} (see however Refs.~\cite{Hayashigaki,Azizi,Wang} for other views). 
These results from QCD sum rules are consistent with the results obtained in analyses using effective chiral models in 
Refs.~\cite{Sasaki:2014asa,Suenaga2} which are based on the chiral partner structure proposed in Ref.~\cite{heavy-partner}. 
At a first glance, an increasing $D$ meson mass with a decreasing chiral parameter however 
seems counterintuitive. Within the naive heavy quark limit, the $D$ meson can be thought of as a heavy quark playing the role of a color source, 
with a constituent quark around it, so that when chiral symmetry is partially restored, the light quark would become lighter making the $D$ meson also light.  

However, as we will see, if confinement persists, the decreasing light quark will allow the light quark to probe larger distances and hence a higher confining energy inducing a 
competition between the kinetic term and the confinement term.   
The combined effect produces a minimum energy value for a certain constituent quark mass so that when the quark mass decreases below the minimum point, 
the mass of the $D$ meson will increase.  In this note, we would like to highlight this effect and discuss why partial chiral symmetry restoration will not necessarily decrease the mass of the heavy-light system. 
We will first try to understand this effect with the help of a simple argument and then confirm the validity of our claim in a realistic and more quantitative constituent quark model calculation. 
  
In the constituent quark model, the energy, or the mass, of a meson made of an anti-charm quark and a light quark can be roughly expressed as
\begin{equation}
E = m_c + m_q + \frac{p^2}{2 m_q} + \sigma r + C \ ,
\label{energy}
\end{equation}
where
$m_c$ and $m_q$ are masses of the charm quark and the light quark, respectively.
$p$ denotes a typical relative momentum between the two quarks, $r$ stands for the typical size of the hadron, and $\sigma$ is the string tension of the potential between the two quarks. 
Note that we omit the Coulomb part of the potential since it will not change the following naive analysis, 
as can be seen in the explicit quark model computation of the next section. $C$ is a constant which is used to fit the total energy to the physical mass spectrum. 
Furthermore, we use $m_q$ as the reduced mass in the kinetic energy term, since $m_c$ is much larger than $m_q$.

\begin{figure}[t!]
  \begin{center}
         \includegraphics[scale=0.8]{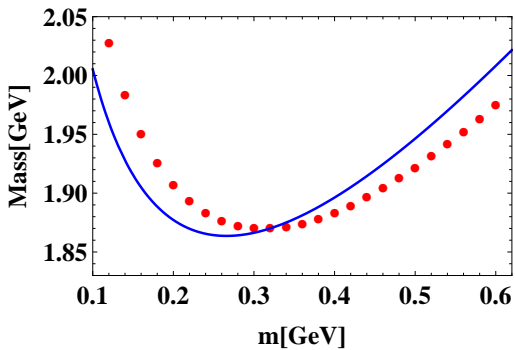}
\caption[]{The solid curve shows the $m_q$-dependence of $E_{\rm min}$ according to Eq.\,(\ref{eq:Emin}) with $m_c+C=0.8$ GeV. 
The red dots represent the $m_q$-dependence of the $D$ meson mass in the full constituent quark model.}
\label{fig1}
  \end{center}
\end{figure}

For a rough estimation of the mass of the heavy-light meson, we first replace $r$ in Eq.~(\ref{energy}) with $1/p$, and minimize the energy with respect to the typical moment $p$. 
The minimization condition provides
\begin{equation}
p = \left( \sigma m_q \right)^{1/3} \ , \label{rel-p}
\end{equation}
which leads to the minimum energy as
\begin{equation}
E_{\rm min} = m_c + m_q + \frac{3}{2} \left( \frac{\sigma^2}{m_q} \right)^{1/3} + C \ . 
\label{eq:Emin}
\end{equation}
Using a conventional value of $\sigma = 0.2\,(\mbox{GeV})^2$, we plot the $m_q$-dependence of $E_{\rm min}$ as a solid curve 
in Fig.~\ref{fig1}. 
As can be observed in the figure, $E_{\rm min}$ takes a minimum value around $m_q \simeq 0.3$\,\mbox{GeV}. 
 
Suppose that $m_q \simeq 0.3$\,\mbox{GeV} in the  vacuum and that it reduces when the chiral symmetry is partially restored.
It is natural to assume that $\sigma$ and $C$ does not change  its value so much as it is related to the confinement effect and changes 
at most by less than 5\% at normal nuclear matter density \cite{Hatsuda:1991ez} or at finite temperatures around $T_c$ \cite{Lee:1989qj}. 
As can be seen in Fig.~\ref{fig1}, the mass of the heavy-light meson hence increases with a decreasing chiral condensate. 
The increase of the mass  for smaller quark masses originates from the increase in the relative separation between the quark, as seen 
in Eq.~(\ref{rel-p}), that leads to a larger contribution to the confining potential in Eq.~(\ref{energy}). This result is consistent with that from 
QCD sum rules \cite{Suzuki:2015est,Hilger}, and from effective chiral models \cite{Sasaki:2014asa,Suenaga2}. 

\section{Constituent quark model calculation}

Let us next perform a more detailed analysis using a non-relativistic constituent quark model.
We start from  the following nonrelativistic Hamiltonian.
\begin{eqnarray}
H=\sum_{i=1}^{n}(m_{i}+\frac{\textbf{p}^2_i}{2m_i})-\frac{3}{4}\sum_{i<j}^{n}\frac{\lambda^c_i}{2}\frac{\lambda^c_j}{2}(V^{C}_{ij}+V^{SS}_{ij}),
\end{eqnarray}
where $m_i$ and $\lambda^c_i/2$ are the quark masses and  the color operators of the $i$'th quark, and  $V^{C}_{ij}$ and $V^{SS}_{ij}$ are 
the confinement and the  hyperfine potential between quark $i$ and $j$, respectively.   We adopt the following form for the confinement and 
hyperfine potential~\cite{Bhaduri:1981pn},
\begin{eqnarray}
V^{C}_{ij}=-\frac{\kappa}{r_{ij}}+\frac{r_{ij}}{a_0}-D, \label{vc_ij-01}
\end{eqnarray}
\begin{eqnarray}
V^{SS}_{ij}=\frac{\hbar^2c^2{\kappa}}{m_im_jc^4}\frac{1}{(r_{0ij})^2r_{ij}}e^{-(r_{ij}/r_{0ij})^2}
{\sigma}_i\cdot{\sigma}_j. \label{v-ss}
\end{eqnarray}
Here, $r_{ij} = |\textbf{r}_i-\textbf{r}_j|$ is the distance between quark $i$ and $j$, and $(r_{0ij})$ is chosen to depend on
the respective masses as follows:
\begin{eqnarray}
r_{0ij}=1/(\alpha+\beta \frac{m_im_j}{m_i+m_j}).
\end{eqnarray}
The parameters are determined to give an overall fit to the meson systems in the light and heavy quark sector \cite{Park:2015nha}. 
It should be noted that including the  pion and sigma exchange potentials is important for a consistent description of three-quark and six-quark states as 
discussed in Refs. \cite{Huang:2013nba,Buchmann:1998mi}. We use the variational method with a single Gaussian trial wave function and determine 
the parameters to reproduce the light-light and light-heavy meson masses.  Using multiple Gaussian functions for the trial wave function leads to negligible 
change to the fitted masses \cite{Park:2013fda}. The resulting parameter values are shown in Table \ref{normalmeson_mass-01}. 

\begin{table}[t!]
\caption{Parameters fitted to the experimental  meson masses using the variational method with a single Gaussian.
The respective units are given in the third row.}
\begin{center}
\begin{tabular}{|c|c|c|c|c|c|c|c|c}
\hline \hline
  $\kappa$  & $a_0$  & $D$ & $\alpha$ &$\beta$ & $m_q$  & $m_c$  & $m_b$   \\
\hline
  0.48  & 5.43  & 0.911  &  2.2    & 0.277        & 0.324  & 1.83 & 5.21 \\
  & $ \mbox{GeV}^{-2}$ & $\mbox{GeV}$   & $(\mbox{fm})^{-1}$  &     & $\mbox{GeV}$   & $\mbox{GeV}$ & $\mbox{GeV}$  \\
\hline \hline
\end{tabular}
\end{center}
\label{normalmeson_mass-01}
\end{table}

\begin{figure}[t!]
  \begin{center}
         \includegraphics[scale=0.8]{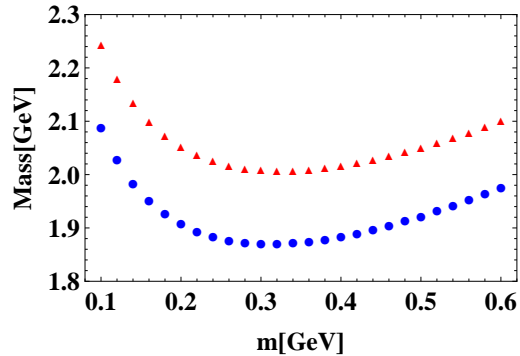}
\caption[]{$m_q$-dependence of $D$ meson (blue circles) and $D^*$ meson (red triangles) masses in the constituent quark model. }
\label{D-meson}
  \end{center}
\end{figure}

\begin{figure}[t!]
  \begin{center}
         \includegraphics[scale=0.8]{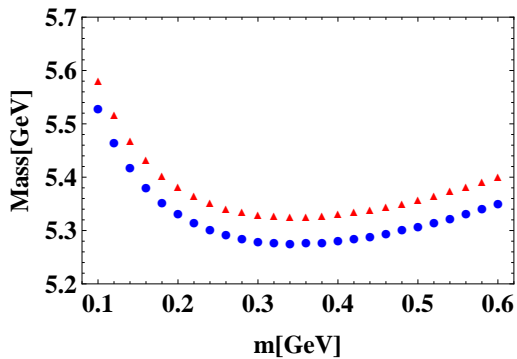}
\caption[]{$m_q$-dependence of $B$ meson (blue circles) and $B^*$ meson (red triangles) masses in the constituent quark model. }
\label{B-meson}
  \end{center}
\end{figure}

The $D$ meson mass results of the full quark model calculation with a changing constituent quark mass are shown in Fig.\,\ref{fig1} as 
red dots. It can be observed in this figure that the behavior of the naive formula of Eq.\,(\ref{eq:Emin}) is qualitatively reproduced by our 
more accurate calculation. 
In Fig.\,\ref{D-meson}, the $D$ meson (blue circles) and $D^*$ mesons (red triangles) masses are given as a function 
of the constituent quark mass, while Fig.\,\ref{B-meson} shows the same for $B$ and $B^*$ mesons. 

As can be seen in Figs.\,\ref{D-meson} and \ref{B-meson}, the full constituent quark model calculation shows that starting 
from the fitted constituent quark mass of 324 MeV for 
the light quark, one finds that for both the $D$ and $B$ mesons, the meson masses increase when the constituent quark mass decreases.  
The main reason for this effect is that with a decreasing constituent quark mass, the kinetic term forces the relative momentum to become smaller and 
hence the light quark to probe larger distances.  This will thus lead to a larger potential energy coming from the string tension.  

\begin{figure}[t!]
  \begin{center}
         \includegraphics[scale=0.8]{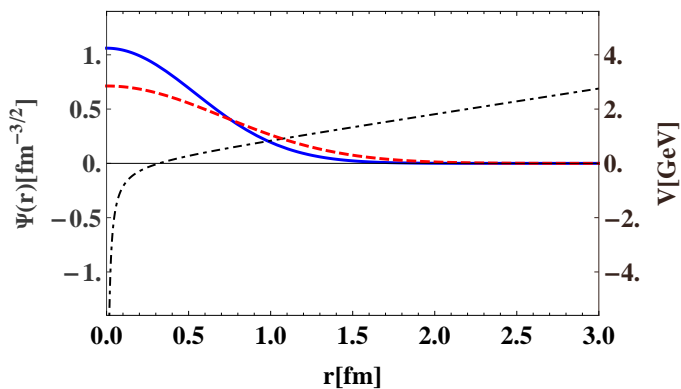}
\caption[]{
 $r$-dependence of the $D$ meson radial wave function 
with $m_q=324$ MeV (blue solid line) and with $m_q=160$ MeV (red dashed line).  
The black dot-dashed line represents the sum of the Coulomb and confining potential. 
}
\label{D-meson2}
  \end{center}
\end{figure}

This effect is well reflected in the wave function obtained for different constituent quark masses.  The blue solid  curve  in Fig.\,\ref{D-meson2} shows the wave 
function obtained with the vacuum constituent mass of 324 MeV, while the red dashed curve  represents that with a mass of 160 MeV; the 
black dot-dashed curve shows the Coulomb and confining part of the potential.  As can be seen in the figure, the wave function for the smaller mass 
spreads out more and thus probes higher values of the potential energy.
Therefore, while a smaller quark mass will decrease the heavy-light meson mass, the contribution from probing larger confinement properties 
will increase its total mass.  

A few comments are in order here.
First, in this simple model we did not introduce any modification of the string tension 
in relation to the (partial) restoration of chiral symmetry.   
This is so because in nuclear matter, while the chiral order parameter is expected to be reduced by more than 20\% at 
normal nuclear matter density, 
the gluon condensate is expected to change by less than 5\%. 
While there is no direct relation between confinement and gluon condensate, some connection can be made between the Wilson loop and gluon condensate 
via the operator product expansion (OPE) \cite{Shifman:1980ui}. Moreover the abrupt (slow)  change of the electric (magnetic) condensate across 
the phase transition can be associated with the critical (soft) change of the space-time (space-space) Wilson loop across the same phase transition temperature \cite{Lee:2008xp}. 
Second, we did not consider any potential $D/ \bar{D}$ splitting in medium. Such an effect could be introduced by including a vector meson exchange type 
interaction of the $D$ mesons with the surrounding medium. Hence, the present mass change 
should be understood as the average mass shift of these mesons. 
Third, we expect that the present framework will break down for too small quark masses as 
the non-relativistic approximation will become invalid.  
Fourth, the present argument will not be reliable for light-light meson systems as the hyperfine potential $V^{SS}$ in 
Eq.~(\ref{v-ss}) will become unrealistically large as the light quark masses decrease. 
$V^{SS}$ in fact will increase at a $1/m_q^2$ rate with a decreasing constituent quark mass, eventually leading to a negative pion mass, which 
does not seem to be realistic.

As a last point, let us discuss the results of the $D^*$ and $B^*$ mesons, shown as red triangles in 
Figs.\,\ref{D-meson} and \ref{B-meson}. It is seen in these figures that these mesons within our model 
behave almost exactly in the same way as their pseudoscalar counterparts. This means that our model predicts 
the $D^*$ and $B^*$ meson masses to increase as chiral symmetry is restored and the constituent 
quark mass decreases. 
The existence of the heavy quark symmetry implies the degeneracy between  pseudscalar and vector heavy-light mesons in the vacuum, 
which is expected to be intact in the dense matter.\footnote{Although two mesons with different spins mix with each other when the spin symmetry 
of light quark sector is broken, the degeneracy between heavy-quark partners persists~\cite{Suenaga1}.} 
Then, the analysis here is consistent with the heavy quark symmetry similarly to the result obtained in Ref.~\cite{Suenaga2}. 
It is not yet clear whether this behavior will be consistent with the predictions of QCD sum rules, 
in which the operator product expansion (OPE) relates various order parameters of chiral symmetry 
to certain properties of the spectral function.   
In the $D^*$ (or $B^*$) channel sum rules, the OPE term involving the quark condensate has the same 
sign as in the $D$ (or $B$) case \cite{Dominguez,Gelhausen,Wang}, which could indicate an increasing mass as chiral 
symmetry is restored, consistent with our model. There are, however, more terms present in the OPE, which could modify 
this naive expectation. 
Therefore, only a full QCD sum rule analysis done in the same way as in Refs.\,\cite{Suzuki:2015est,Hilger} will make a 
consistency check with our model possible. We leave this topic for future work. 


\section{Summary}

In this work, we have studied masses of heavy-light meson in a dense environment such as nuclear matter, in which chiral symmetry is partially 
restored. For this purpose, we have made use of a simple quark model with one heavy and one constituent light quark. This constituent quark 
becomes lighter as chiral symmetry is restored and therefore likely changes the mass of the whole heavy-light meson. 
Naively, one would expect the heavy-light meson to decrease its mass as one of its constituents becomes lighter. 
We have however shown in this note that this is not necessarily the case. 
Assuming that the confining potential remains approximately constant at normal nuclear matter density, 
we have demonstrated that the wave function spreads out as the constituent quark mass decreases and therefore 
receives a larger potential energy due to the linearly rising confining potential. 
For constituent quark masses below about $0.3$ GeV, this effect leads to an increase of the heavy-light meson, 
as shown for instance in Fig.\,\ref{fig1}. 
In our quark model calculations, we have examined states containing both a charm and bottom quarks and found 
the same behavior for both cases (see Figs.\,\ref{D-meson} and \ref{B-meson}). 
Our results for $D$ and $B$ mesons are consistent with recent sum rule analyses which obtain increasing masses with 
increasing density \cite{Suzuki:2015est,Hilger} and with calculations based on the Skyrme crystal model \cite{Suenaga2}.  
For the future, it will be interesting to check whether the predictions of our model for the $D^*$ and $B^*$ mesons 
can be confirmed and reproduced by these more QCD-related  approaches.

\textit{Acknowledgements} 
SHL AP and WP were supported by the Korea
National Research Foundation under the grant number
KRF-2011-0020333 and KRF-2011-0030621. 
The work of MH was 
supported in part by the JSPS Grant-in-
Aid for Scientific Research (C) No. 24540266. 
The work of CN is supported in part by the JSPS Grant-in-
Aid for Scientific Research (S) No. 26220707. 
MH  and CN are grateful for the hospitality of the members of the Institute of Physics and Applied Physics during their stay in 
Yonsei University where the main part of this work was done.

\end{document}